\documentclass[man,apacite,floatsintext]{apa6}

\usepackage[english]{babel}
\usepackage[utf8x]{inputenc}
\usepackage{float}
\usepackage{amsmath}
\usepackage{amsfonts}
\usepackage{graphicx}
\usepackage{svg}
\usepackage[colorinlistoftodos]{todonotes}
\restylefloat{table}

\title{Understanding Our People at Scale with Machine-assisted Psychoanalysis}
\shorttitle{Understanding Our People at Scale}
\author{Tam Nguyen}
\affiliation{The North Carolina State University}
\authornote{
Paper for the ********* Conference.

Please come back for correct citation of this work.

The author has no conflict of interest to disclose.

Correspondence concerning this paper should be addressed to tam.nguyen@ncsu.edu

https://www.linkedin.com/in/theresearchninja/

}

\abstract{Human psychology plays an important role in organizational performance. However, understanding our employees is a difficult task due to issues such as psychological complexities, unpredictable dynamics, and the lack of data. Leveraging evidence-based psychology knowledge, this paper proposes a hybrid machine learning plus ontology-based reasoning system for detecting human psychological artifacts at scale. This unique architecture provides a balance between system's processing speed and explainability. System outputs can be further consumed by graph science and/or model management system for optimizing business processes, understanding team dynamics, predicting insider threats, managing talents, and beyond.}

\begin{document}
\maketitle

\section{Introduction}
There is a strong relationship between individual psychology and organizational performance. Leaders respond to organizational developments and challenges with partial reliance on their own knowledge, past experiences, and most importantly, their personalities \cite{Hambrick2007UpperUpdate}. \citeA{Berson2014LeadingMotivation} identified that CEO's personal values are strong predictors of organizational culture and performance. It could be explained by the attraction-selection-attrition perspective \cite{Schneider1995TheUpdate}, from which a leader is able to attract certain types of followers per his/her own interesting personality, and the followers are more open to express as well as adopt new ideas \cite{Oreg2011LeadershipStyle}.

Unethical decision makers are referred to as corporate psychopaths \cite{Stevens2012SuccessfulWhy} although machiavellians and narcissists are also included \cite{Furnham2013TheReview.}. Their manipulative nature and charming appearances may get them through job interviews, to high ranks with broad influences. The negative consequences of having them in charge may include: radical organizational changes leading to instability \cite{Chatterjee2007ItsPerformance.}; higher psychological exhaustion and lower job satisfaction scores \cite{Volmer2017TheCorrigendum.}; employees' higher moral disengagement \cite{Christian2014TheWork} and lesser appetite to take smart risks \cite{Liu2016AbusiveIdentification.}. On a global scale, corporate psychopaths could be a reason behind the recent global financial crisis \cite{Boddy2011TheCrisis} as they are more willing to risk other people's money for their own personal gains \cite{Jones2014RiskTriad.}, and in some cases, commit frauds \cite{Perri2013VisionariesProphets}.

Steve Jobs was a corporate psychopath, and he was fired from the company he founded \cite{Isaacson2013SteveBiography}. However, John Sculley - the Apple CEO who fired Steve Jobs - admitted that the firing was his own biggest mistake \cite{NBC2011JohnJobs}. Indeed, most of Steve Jobs' psychopathic behaviors were permanent, and the only difference between his firing and his then legendary comeback was the deeper awareness of both organizational and inter-personal issues, consequently leading to better informed behaviors and team dynamics. The firing's role as a necessary shock for better changes in both Steve Jobs and Apple is debatable. On one hand, it could be the only way to force Steve Jobs to realize the negative consequences of some of his behaviors. On the other hand, it could be avoided, and the amazing Apple products could have arrived several years sooner. Ultimately, reliable data and well constructed evidences would be needed in order to settle a debate like this one.

Therefore, this paper proposes a system for large-scale, constant mining of psychological artifacts from texts, leveraging evidence-based clinical psychology knowledge, natural language processing (NLP), machine learning (ML), and ontology-based reasoning. First, a manual psychoanalysis of Steve Jobs' bibliography demonstrates the process' common steps. Second, the paper presents a high-level system architecture, in which manual steps were conceptualized into stacks, their components and corresponding relationships. Mechanisms crucial to the system's core capabilities will be described. Third, development details involving related works, technology stacks, and development practices will be recommended. Finally, the paper discusses further integration of the system in optimizing business processes with graph science and model querying, and in understanding bad cyber actors. Other issues such as legality and privacy will also be discussed.

\section{The Case of Steve Jobs}

Diagnosed with cancer, Steve Jobs allowed Isaacson complete freedom in aggregating information from both Steve's friends and foes. "Steve Jobs: The Exclusive Biography" \cite{Isaacson2013SteveBiography} was released in 2012, offering insights to both amazing and terrible events in Steve Jobs' life. From this biography, the paper diagnosed Steve Jobs with Obsessive Compulsive Personality Disorder (301.4) and Narcissistic Personality Disorder (301.81). Figure 1 and 2 respectively represent the DSM 5 \cite{APA2014DiagnosticDSM5} dimensional scores.  This manual process can be summarized as followed. First, a psychologist will carefully read the written texts. Second, relevant details will be extracted, grouped, and assigned an observation codes (o.code). Third, o.codes will be matched with DSM 5 diagnostic criteria. An o.code may score in different diagnostic criteria, across different disorders. A dimensional score is the sum of all the matched o.code(s) for that particular dimension. Finally, conclusion(s) will be drawn from the dimensional scores, following previous evidence-based research results such as the DSM 5. This process is based on forensic psychoanalysis \cite{Chadda2013ForensicPsychiatry}, of which results are acceptable within the court of laws. Further details will be explained below.

\begin{figure}[t]
\centering
\begin{minipage} {0.45\textwidth}
    \centering
    \includegraphics[width=1\linewidth]{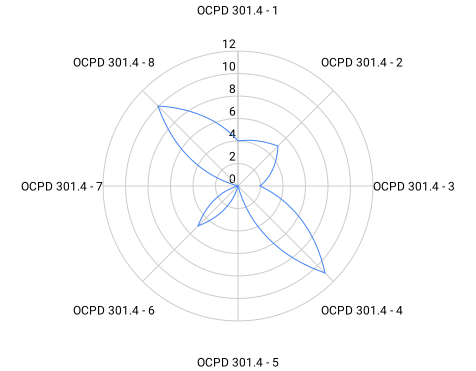}
    \caption{Obsessive Compulsive Personality Disorder Profile}
\end{minipage} 
\begin{minipage} {0.45\textwidth}
    \centering
    \includegraphics[width=1\linewidth]{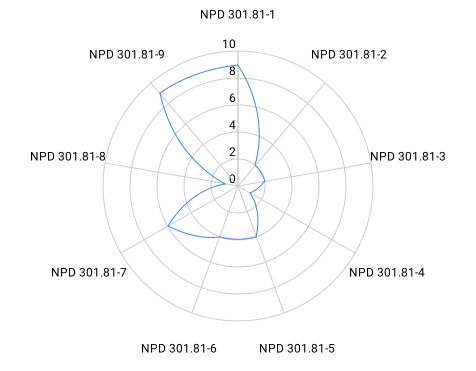}
    \caption{Narcissistic Personality Disorder Profile}
\end{minipage}%
\end{figure}

There are at least 35 evidences, grouped into 7 groups. Each evidence inside a group will have an observation code in the format of the group number to be followed by the observation number. Details regarding group O2 and O4 will not be listed here, as they are not relevant to the diagnosis presented in Table 1.

\subsubsection{O1 - Steve had a strong desire for complete control} Del Yocam believes that it has something with Steve's at-birth abandonment (O1-1) \cite{Isaacson2013SteveBiography}. Andy Hertzfeld - a close friend of Steve - reported that Steve saw products and environments as extensions of himself which he must gain complete control (O1-2), yet he could not control himself from being cruel and harmful to other people (O2-1) \cite{Isaacson2013SteveBiography}. Steve believed users should not be able to extend Apple I, even for their own use (O1-3). Steve can completely shut unwanted interference out, even with sensitive personal matters such as the Brennan's pregnancy (O1-4) \cite{PsychiatricTrimes2014TheGeniuses}. Steve wanted Apple's computer case can only be opened by Apple's special tools (O1-5). 

\subsubsection{O3 - Steve believed he is special and was willing prove it} He was eager to do all the extra problems that his math teacher gave him (O3-1). In class, he would sat in a corner, doing his own things (O3-2). In college, he was bored of required classes, dropped out, and audited the classes he linked (O3-3). At Apple, he insisted to be placed ahead of Woz and refused to be employee number 2. He later claimed number zero for his company badge even though he was still number 2 as payroll system started with number 1 (O3-3). Steve preferred people jump when he said "jump" (O3-4) \cite{Isaacson2013SteveBiography}. Hertzfeld recalled that Steve did believe he was the chosen one, on par with Einstein and Gandhi (O3-4) \cite{Isaacson2013SteveBiography}. When negotiating with Microsoft for the first time, Steve behaved in a way as if he was allowing Bill to be a part of the collaboration (O3-5). Redse - the girlfriend who Steve loved the most - acknowledged that Steve had been very self centered, and that it was his job to teach people aesthetics (O3-6) \cite{Isaacson2013SteveBiography}.

\subsubsection{O5 - Steve uses drugs} Steve was known to use marijuana, and LSD (O5-1) \cite{Isaacson2013SteveBiography}. He regarded LSD as an important part of his life as it helped him discover the other side of his consciousness during meditation sessions (O5-2). 

\subsubsection{O6 - Steve can go to the extreme in the ways he does things} Friedland reported that Steve can carry things to the extreme with high intensity at times (O6-1). When choosing the color for the Apple computer plastic case, none of Pantone's two thousand shades of beige was good enough for Steve (O6-2) \cite{Isaacson2013SteveBiography}. Steve decided to go with almost no furniture in his house because no furniture fit his high standards (O6-3). Atkinson recalled that Steve cannot make trade-off (O6-4). Steve was not able to tolerate a tiny mold line in the chassis of his NEXT computer and forced the manufacturer to buy a \$150,000 sand machine to solve the problem (O6-5). When building NEXT factory, Steve demanded the walls to be painted white, certain robots to be painted with certain colors, and putting \$20,000 leather chairs in (O6-6). 

\subsubsection{O7 - Steve carried some stubborn and/or weird beliefs} He believed that his strict vegetarian diet will not cause body odor which is not true, and he would not use deodorant nor shower regularly (O7-1) \cite{Isaacson2013SteveBiography}. Even when paternity test confirmed he was the father of Lisa, he kept telling others that he was not (07-2). Bill Atkinson said Steve had the ability to deceive even himself (O7-3) \cite{Isaacson2013SteveBiography}. Steve has a binary world view - everything is either "excellent" or "total shit" to him (O7-4).

\begin{table}[t]
\centering
\resizebox{0.9\columnwidth}{!}{
\begin{tabular}{|l|l|l|l|l|l|l|l|l|}
\hline
\textbf{O codes} & \textbf{301.4 - 1} & \textbf{301.4 - 2} & \textbf{301.4 - 3} & \textbf{301.4 - 4} & \textbf{301.4 - 5} & \textbf{301.4 - 6} & \textbf{301.4 - 7} & \textbf{301.4 - 8} \\ \hline
O1-1 &  &  &  &  &  & 1 &  &  \\ \hline
O1-2 &  &  &  &  &  & 1 &  &  \\ \hline
O1-3 &  &  &  & 1 &  &  &  & 1 \\ \hline
O1-4 &  &  & 1 &  &  &  &  &  \\ \hline
O1-5 &  &  &  & 1 &  & 1 &  &  \\ \hline
O3-2 &  &  &  &  &  &  &  & 1 \\ \hline
O3-3 &  &  &  &  &  &  &  & 1 \\ \hline
O3-3 & 1 &  &  &  &  &  &  & 1 \\ \hline
O3-6 &  &  &  & 1 &  &  &  &  \\ \hline
O5-2 &  &  &  & 1 &  &  &  &  \\ \hline
O6-1 &  & 1 & 1 &  &  &  &  &  \\ \hline
O6-2 & 1 & 1 &  &  &  &  &  & 1 \\ \hline
O6-3 & 1 & 1 &  & 1 &  &  &  &  \\ \hline
O6-4 &  & 1 &  &  &  & 1 &  & 1 \\ \hline
O6-5 &  & 1 &  & 1 &  &  &  &  \\ \hline
O6-6 & 1 &  &  & 1 &  & 1 &  & 1 \\ \hline
O7-1 &  &  &  & 1 &  &  &  & 1 \\ \hline
O7-2 &  &  &  & 1 &  &  &  & 1 \\ \hline
O7-3 &  &  &  & 1 &  &  &  &  \\ \hline
O7-4 &  &  &  & 1 &  &  &  & 1 \\ \hline
\end{tabular}
}
\caption{Diagnosis of Obsessive Compulsive Personality Disorder}
\label{tab:OCPD}
\end{table}

Table 1 illustrates the mapping of o.codes to respective DSM-5 diagnostic criteria \cite{APA2014DiagnosticDSM5} for Obsessive Compulsive Personality Disorder (OCPD). This diagnosis is the key in opening up pathways for figuring out the strengths and weaknesses of an OCPD person, and how to translate those findings into better team performance. For example, research has shown that OCPD people have enhanced visual performance \cite{Ansari2016EnhancedDisorder}, and difficulties with interpersonal functioning \cite{Cain2015InterpersonalDisorder}. A better understanding of issues will lead to better identification of evidence-based interventions such as cognitive behavioral therapy, psychodynamic therapy, dialectical behavior therapy, group-based therapy, and so on \cite{Alex2010PsychologicalDisorder}.

\section{The Machine-assisted Psychoanalysis System}
The Machine-assisted Psycho-Analysis System (MPAS) puts the above-mentioned process into automation, aiming at continuous mining of psychological artifacts that can be further analyzed for better situational awareness (short term) and strategical plannings (long term). At the highest level, MPAS consists of a Note stack, a Card stack and the Interface stack as shown in Figure 3. 

\begin{figure}[t]
\centering
\includegraphics[width=5.5in]{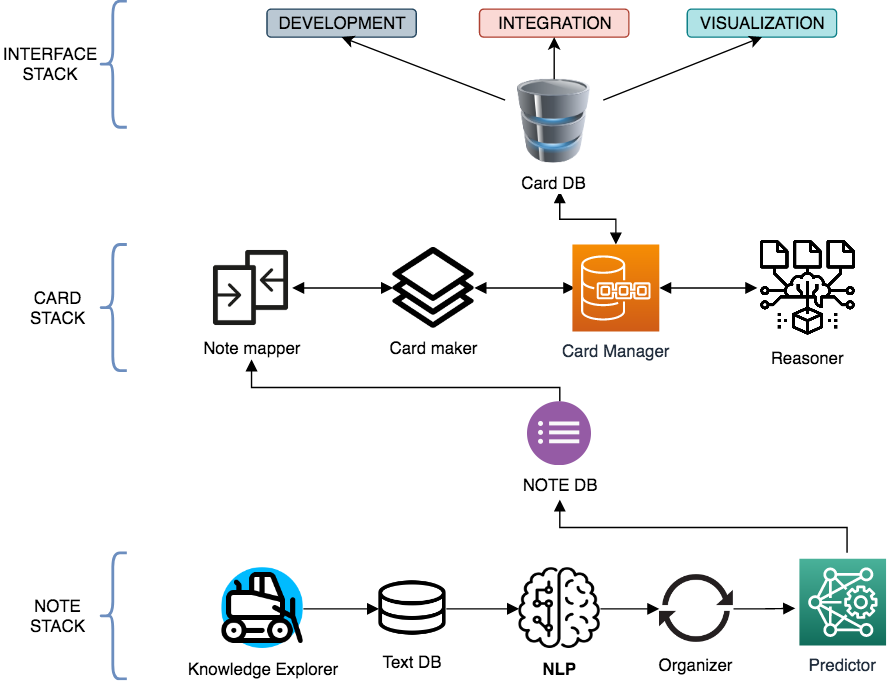}
\caption{MPAS architecture}
\end{figure}

\subsubsection{The Note stack} It is at the lowest level, responsible for creating "Notes" which are the observations in the manual psychoanalysis of Steve Jobs. At the beginning, the Knowledge Explorer component automatically crawls identified data sources, which may include straight-forward text contents (emails, forum posts, survey feedback), documents in certain formats (MS Word, Adobe PDF), packaged files (zip, rar), and other less common data sources (program logs, speech-to-text, etc). Hence, the crawler must execute pre-processing tasks such as extracting raw texts from those data sources, and assigning meta-data. Time, place, main subjects, and links to original sources are crucial meta-data. The Text DB stores all of those data and meta-data for further processing by the NLP component.

The Natural Language Processing (NLP) component has within itself an ontology, a set of dictionaries, and a machine-learning (ML) model. The ontology describes the Entities and their Relationships within a certain knowledge domain. For example, the "Alcohol consumption" knowledge domain may include main entities of: drinkers, liquor, wine, beer, amounts; and main relationships of: consume, purchase, and discuss. The dictionaries document the entities and relationships' polymorphism. Different liquor brand-names are polymorphisms of the entity "liquor", as "drink", "booze up", "bottom up" are those of the "consume" relationship. Based on previous training, the ML model will annotate raw chunks of texts with recognized entities and their relationships. Of course, the NLP will also look for further details regarding subjects, time and place. The Organizer will then organize chunks of texts by main subjects, associated place and time. Alignment of variables such as Time can be crucial in understanding certain behaviors such as alcohol addiction, and recognizing duplicated contents. The Organizer operates on its own discrete database.

The Predictor is another trained ML model responsible for making "Notes", each of which summarizes several chunks of annotated texts. For example, in making a note about a person A (subject), the Predictor does not see all the human texts but only marked entities and their relationships. If it sees A consumes alcohol several days a week for several weeks then it may produce a note such as "Person: A; Action: consumes alcohol; Intensity: very frequent; Confidence: high; time and place information; tracking information". We may consider the entities and relationships as inputs while a note is the output, and the ML model can be trained on pre-identified input-output pairs (ground truths). All notes will be saved into the Note database (Note DB).

Notes can be really noisy due to several reasons. The ML models are not perfect and will make wrong prediction and/or overlook data. This is especially true when human writings are unpredictable and imperfect. People sometimes lie or exaggerate. Synchronizing data is another big challenge as there must be a reasonable window for the Organizer to hand off data to the Predictor. Some facts may fall into one processing window while others may fall to another, resulting in two notes about a same certain event or period. Speed rather than accuracy is the main focus of this stack.

\subsubsection{The Card stack} It is strictly about logical inferences in order to self-correct the accumulated noises being passed up from the Note stack. First, the Note mapper reads new data from Note DB. The read schedule can be pre-set or be triggered by the Card Manager. The Note mapper will check if Notes are new and in proper format. Following a pre-designed ontology, the Note mapper will refine the acquired notes following: a maximum value rule (for example, in dealing with differences in the reported amounts of beers A had drank in a particular party); a combining value rule (for example, in assessing A's monthly alcohol consumption rate by using various notes about weekly alcohol consumption); a majority rule (for example, in addressing conflicting notes). Rules can be applied differently according to the data (entities and relationships) in the  note fields.

The Card Maker grabs the refined notes and maps them to proper cards. This action is similar to assigning o.codes to the DSM 5 dimensions of each potential disorder. Each card is a collection of notes and was structured per the pre-defined ontology. For example, the DSM5-based ontology for Obsessive Compulsive Personality Disorder consists of 8 dimensions (the columns in Table 1), and the disorder can be diagnosed when 4 dimensions (criteria) are met. In this case, a card will be made when there are enough refined notes to fill at least 4 dimensions, and the Card Manager will be notified. Otherwise, the Card Maker holds on to the pre-mature cards, awaiting further refined notes.

In the mean time, the Card Manager will work with the Reasoner in order to identify stealthy inconsistencies among new cards, and/or between new cards and old cards. Naturally, people and things change. For example, when a difficult project is over, a manager may become free of anxiety symptoms. In that context, the conflict between a new card (saying about up-beat attitudes) and a previously committed one (saying about anxiety) can be resolved by putting an expiration date on the old card. In the bi-polar context, conflicting information about moods may actually indicate mood-swings. Therefore, the Reasoner must use and only use evidence-based logics to infer the right decision. The DSM 5 \cite{APA2014DiagnosticDSM5} describes such logics in great details. Conflicts among new cards such as identification of mental issues that were clinically proven to be not comorbid maybe due to errors at the Note stack, or purposeful lying.

We should note that the relationships among Card stack's components are bi-directional while those at the Note stack are one-directional. From a data read-write perspective, Note stack components have read-down/write-up relationships, while members of the Card stack can "talk" to each other at the same level. Although, only the Card Manager may commit cards to the official Card DB. For example, upon the Reasoner's detection of conflicts between card X and card Y, the Card Manager may initiate card remake request(s) for either X or Y or both. Card Maker will check with the Note Mapper to see if there is any additional note to be added. A waiting period may be specified by the Card Manager in order for more new notes to be passed up from the Note stack.

\subsubsection{The Interface stack} It is at the highest level starting with the Card DB, which can be consumed by the Development module, the Integration module, and the Visualization module. The Visualization module provides options to view networked cards, drill down to cards' lower level data, and ideally, to visually simulate different scenarios. Simulating scenarios on a networked graph is similar to finding all possible routes between a start point and an end point on a map. Integration module allows exchanges of knowledge accross different domains. For example, psychiatric cards can be analyzed in tandem with cyber security threat cards in order to infer cyber risks. The Discussion section will go into further integration details. Lastly, the Development module represents all other development efforts on top of the Card DB. Unlike integration and visualization module development, these developments may involve other departments, third-parties, contractors and hence, be put under stricter controls.

\section{Recommendations for System Development}
The paper will not provide a pre-made open-source solution due to various reasons including but not limiting to: issues with actual data used for model training (such data must come from the same organization/environment to which the final solution is deployed), psychology is a broad field and it must be tailored to specific business needs from the beginning. However, this section will provide the necessary resources for building up a really good first step.

\subsection{Related Works}
There are three common approaches in identifying human behaviors. Data-driven approach relies on statistical methods and machine learning (ML) techniques to detect patterns. For example, statistical feature quality group selection was used to recognize daily living activities \cite{Banos2012DailySelection}, and support vector machine was used to recognize basic speech emotions \cite{Seehapoch2013SpeechMachines}. On the other hand, the Knowledge-driven approach uses ontologies and rules. For example, a rule-based multi-modal approach is also able to identify human emotions \cite{Khanna2013RuleApproach}, as well as human basic behaviors \cite{Bae2011AnHomes}. Compared to the Knowledge-driven approach, the Data-riven approach has two fundamental differences. First, it requires enough samples in order to perform accurately, and there is no exact science on how to calculate what enough is enough. Second, domain specific intuitions and explain-ability are not completely supported. For example, ML models cannot explain their own conclusions. The Knowledge-driven approach also has its own drawbacks. It performs slower and its ontology can be really difficult to build. The biggest issue with ontology design is the lack of gold standards or ground truths, leading to issues such as defining the property range and scope of classes. Those pros and cons gave birth to the Hybrid approach, combining both Data-driven and Knowledge-driven techniques. For example, ontology-based stream reasoning helps with continuous activity recognition \cite{BakhshandehAbkenar2014MyActivity:Reasoning}, and complex human activities can be recognized using OWL 2 modeling and reasoning \cite{Riboni2011OWLActivities}. This paper was inspired by the work of \cite{Villalonga2016Ontology-basedIdentification}, in which the architecture was separated into lower and higher levels. Different types of primitives were gathered at the lower level, only to be formed into "contexts" at the higher level. 

In the context of understanding team behaviors through the lens of personal psychology, there is an emerging array of technologies. \citeA{Stevens2012CognitiveTeamwork} used cognitive neurophysiologic synchronies, which are low-level derivatives of electroencephalography (EEG) measurements, to identify correlations between team cognition and changes in members' speech. The results can only be used as complimentary metrics to other methods. \citeA{Gorman2012MeasuringApproach} used nonlinear dynamics and real-time communication pattern analysis to analyse team dynamics and found that intact teams' determinism can increase over missions while it was not the case for mixing teams. This proves the usefulness of communication analysis over other noisy channels such as EEG. \citeA{Salas2015TeamsEffectiveness} applied latent-semantic-analysis on text-based data sources such as chat logs, blogs and journals, and was able to assess aspects of team dynamics through identified basic emotional states such as anxiety, being comfortable, and being routine. These methods carry some fundamental limitations such as: mandatory sensor wearing, interruption of work flow (in order to fill out probing questions), and difficult integration with other systems (emotional states are vague and cannot be re-used outside of its native context). On the contrary, the proposed MPAS architecture supports endless integration based on ontologies and logical reasoning. 

\subsection{Ontology building}
There are three main steps to engineer ontologies from scratch \cite{Uschold1995TowardsOntologies, Uschold1995TheOntology}: 1. Identifying the purpose by examination of motivationg scenarios; 2. Capturing the concepts and their relationships; 3. Codifying the ontology using ontology languages. Knowledge process is the management of knowledge such as the mapping of a domain's knowledge to an ontology \cite{Schreiber2018KnowledgeManagement}. In the domain of clinical psychology, evidence-based knowledge has already been organized into diagnostic manuals and assessment repositories. The recommended resources are: the Diagnositc and Statistical Manual of Mental Disorders \cite{APA2014DiagnosticDSM5}, the Mental Health Intake Evaluation forms \cite{APADiv.31MentalForms}, the APA Online Assessment Measures \cite{APA2013OnlineMeasures}, the Mental Status Examination guide \cite{Norris2016TheExamination}. There is also a large number of more niche assessments. For example, in studying the relationships between leader psychopathy and organizational deviance, the Dirty Dozen scale (psychopathy assessment) was used in conjunction with the organizational deviance scale and the 24-point moral disengagement scale \cite{Erkutlu2019LeaderDisengagement}. Existing ontologies such as the Mining Minds Context Ontology are publicly available for integration and development \cite{Villalonga2016Ontology-basedIdentification}. A list of clinical reasoning ontologies is also available in the work of \cite{Dissanayake2019UsingSynthesis}. Ontologies can also be automatically extracted \cite{Maedche2003ManagingWeb, Haase2010ConsistentOntologies} with existing tools such as TextToOnto and Text2Onto \cite{Cimiano2010Text2Onto}. Other techniques include the application of statistical methods to the Web \cite{Turney2002MiningTOEFL}, and the calculation of semantic relatedness with respect to a taxonomy or semantic network such as WordNet \cite{Resnik1998SemanticLanguage}. 
 
In combining existing ontologies or developing new ones, Content Ontology Design Pattern can facilitate the combination, merging and alignment \cite{Gangemi2005OntologyContent} of ontologies. There are 6 general ontology design patterns: Structural, Correspondence, Content, Reasoning, Presentations, and Lexico-Syntactic \cite{Presutti2008AOntologies}. Last but not least, we can always evaluate an ontology by its: 1. Accuracy (as in representing the domain of interest); 2. Adaptability (as in allowing extension/modification); 3. Clarity (as in defining terms); 4. Cognitive adequacy (in translating to cognitive semantics); 5. Completeness; 6. Conciseness; 7. Expressiveness; 8. Grounding (as in requiring less assumption).

\subsection{The Recommended Technology Stack}

\begin{itemize}
\item Databases: NoSQL \cite{GeeksforGeeks2018IntroductionNoSQL} for Text DB due to heavy write operations. Traditional SQL for Note DB do to heavy look-up operations, providing that data schema of the Predictor's outputs is fairly stable. Graph database \cite{Neo4jNeo4jDatabases} for Card DB in order to maximize compatibility with further graph-based analysis.
\item Knowledge Explorer: Any already available open-source or commercial solutions. 
\item Natural Language Processing: Stanford core NLP \cite{StanfordUniversityCoreNLPSoftware} or IBM Watson for NLP \cite{IBMWatsonStudio}.
\item Organizer: Own modified version of Mining Mind data curation layer  \cite{UbiquitousComputingLab2019Ubiquitous-computing-lab/Mining-Minds}
\item Predictor: Any already available open-source or commercial machine learning solutions.
\item Note mapper and Card Maker: Own modified version of Mining Mind information curation layer \cite{UbiquitousComputingLab2019Ubiquitous-computing-lab/Mining-Minds}
\item Card Manager: Own modified version of Mining Mind knowledge curation layer \cite{UbiquitousComputingLab2019Ubiquitous-computing-lab/Mining-Minds}
\item Reasoner: Pellet \cite{SirinPellet:Reasoner} or any already available open-source or commercial solutions.
\end{itemize}

\section{Discussions}

\subsubsection{In optimizing business processes}
\cite{Filho2019TeamEfficacy} recently proposed a systemic theory for understanding and explaining team dynamics. This Team Dynamics Theory (TDT) addresses the needs for a parsimonious yet integrated theory, and is a generative nomological network of team cohesion (CO), mental models (TMM), coordination (CD), collective efficacy (CE), and team outcomes (TO). Each of the component can be further broken down, and be mapped from prior research works. For example, prior research works have shown that: team cohesion has a sub-component which relates to individual and team resilience; team cohesion features are different among teams and fluctuate greatly; last but not least, there is a need for further investigation on team resilience constructs \cite{Salas2015TeamsEffectiveness}.

\begin{figure}[t]
\centering
\includegraphics[width=3.5in]{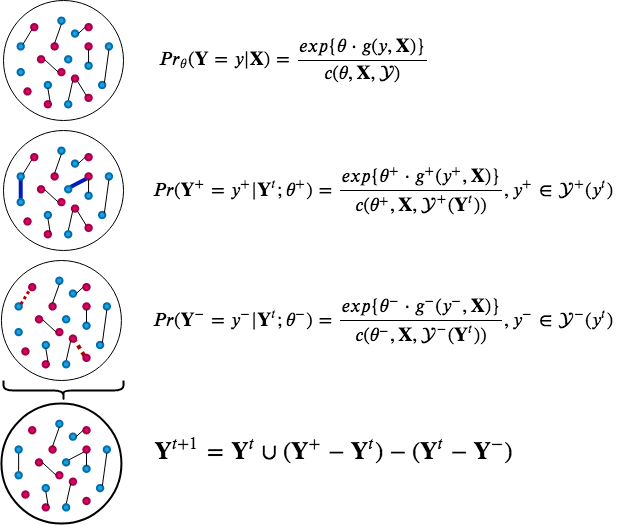}
\caption{STERGM Demonstration}
\end{figure}

The MPAS results can be seamlessly integrated to the TMM component of TDT, and can be confirmed by \cite{Filho2019TeamEfficacy} via two key points. First, TDT affirms the consensus that explicit and implicit communication exchanges are necessary for constructing mental models, which are endowed mechanisms of cognitive-affective-behavioral information. MPAS does process communication texts, and was purposefully designed to recognize both explicit information (at the Note layer) and implicit information (at the Card layer). Second, TMM always stands in relationships with others in the TDT chain, and it evolves dynamically over time. For example, TMM can operate as formative indicators of CD, and covers both team members' communal as well as idiosyncratic knowledge.Because MPAS' cards are ontology-based, their entities and relationships can be easily passed down the logic inference chains, together entities and relationships from other TDT modules. Specifically, if the same architecture (the MPAS architecture) is used, we will end up with CO cards, TMM cards, CD cards, CE cards and TO cards; which in turn, can all be utilized to infer higher level knowledge relating to team dynamics. Those relationships and entities can also be easily mapped out for further analysis leveraging network science methodologies. 

The maps of cards and their relationships are called "networks". Their structural patterns such as reciprocity and temporal patterns can give interesting insights into team dynamics \cite{Antone2020TestingControls}. Separable Temporal Exponential Random Graph Models (STERGMs) \cite{Krivitsky2019STERGM-SeparableStatnet} is one among the latest solutions for investigating network dynamics. Specifically, if we have $Y$ as the random variable for the state of our network with a realization $y$ (ties); the set of possible ties $\mathcal{Y}\subseteq{2^Y}$ where $\mathbb{Y}\subseteq\{1,...,n\}^2$ is the set of potential relations; $g(y)$ is the vector of model statistics, $\theta$ is the coefficient vector, $t$ is a certain point in time, $Y^+$ is the formation network, and $Y^-$ is the dissolution network; then we may predict progression of $Y$ through time as shown in Figure 4. In this case, such insights may explain phenomenons like: individual stress level is higher than overall team stress level, job performance of certain team members are high but team performance is low, and last but not least, an individual performs differently in seemingly identical teams. 

High-level inferences from cards can be quantified with higher precision than machine-learning based results and faster speeds than human experts. Consequently, mental state scores may inform better business decisions such as adapting business processes. A large business may maintain a database of business process models, which in some cases can be at the thousands or even the hundred-thousands \cite{Kunze2015QueryingInclusion}. A decision maker may need to find business process models similar to a given one for reasons such as: assessing potential risks in different scenarios, finding alternatives or optimizing the existing one, and merging departments or companies. \citeA{Kunze2015QueryingInclusion} proposed a querying method that is based on trace semantics of process behavior, abstract trace inclusion, and ranking of query results. Figure 5 presents basic building blocks of a business process model, of which different combinations lead to different traces. \citeA{Kunze2015QueryingInclusion} emphasized on the ranking of behavioral distances as the methodology's key differentiator, which the paper completely agreed on. However, behavioral distances should include both business behaviors (tactics to achieve certain business goals) and mental behaviors (team mental model scores at the least). For example, there are different ways for a department to archive the same business goals, and we need to watch out for the ones that may cause unnecessary psychological stresses (short term or long term) on the teams. 

\begin{figure}[t]
\centering
\includegraphics[width=6in]{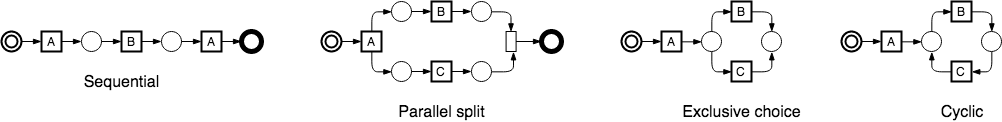}
\caption{Net System Patterns}
\end{figure}

\subsubsection{In understanding bad actors} White-collar criminals are complex in nature. Some even compare them to chameleons, reflecting their superb capability to adapt and manipulate. At personal level, previous works have identified risk factors such as: high scores of narcissism and hedonism; low scores of conscientiousness and self-control; high dimension scores of antisocial personality disorder; as well as certain score variations in extroversion, agreeableness, conceitedness, neuroticism, intellectualism, and negative valency \cite{Brady2016ForensicCrime}. To a certain degree, white-collar criminals share some psychological patterns with other criminals. While most of them may commit crimes for financial gains, some of them commit white-collar crimes for other reasons, including cold-blooded calculations. This landscape only gets more complicated at the institutional level. For example, a senior manager with high scores on risk factors may remotely instruct an employee with high loyalty score (to that manager) to commit white-collar crimes. Other complications may include: shared feelings, defense mechanisms, fantasies, beliefs, the tendency to give superiors the benefit of the doubt, traumatic avoidance, denial and cult-like shared or enabled delusions \cite{Brady2016ForensicCrime}. On this note, the paper cannot help but emphasize the importance of looking at psychological artifacts in a networked view, and paying attentions to the dynamics among the components. While MPAS advocates for using psychology forensic science in engineering its ontologies which are the main component of automatic reasoning, it is important to note that psychology forensic science is far from perfect \cite{Brady2016ForensicCrime}. MPAS can be used internally, externally or both; and for identifying mental risk factors, as well as their combinations in higher-level analysis. Leveraging MPAS for risk assessment requires further steps of quantifying the risk factors, simulating scenarios, and producing risk estimates.

\subsubsection{On potential issues}
There are many issues around Artificial Intelligence (AI) and Machine Learning (ML) systems. The biggest challenge concerns the legal definitions of AI and ML. Even the father of AI could not define what is AI \cite{Scherer2016REGULATINGSTRATEGIES}. This leads to the issues of defining the "Actor", which is the target for prosecution in court cases. MPAS clarifies this issue via its layered architecture. ML can be fully deployed but is confined within the Note layer. This minimizes the negative impacts of ML's mistakes. System designers and operators will have more controls at the Card layer where they can design the logics which card maker and manager rely on. Human bias should be at minimal if the logics are based on strong research results. Unlike other ML-based solutions which operate like black-boxes \cite{AnjomshoaeExplainableReview}, MPAS hybrid architecture of ML and human-controlled logic reasoning provides the system users/administrators with explainable findings, and makes them accountable for the final decisions. In this case, explain-ability is represented by recorded reasoning chains stored in cards, and the ability to trace back to the notes that formed a card.

Workplace surveillance versus employees' privacy is another concern. It is important to note that User Activity Monitoring (UAM) - a subset of workplace surveillance - already has a market of \$1.1 billion and is expected to grow to \$3.3 billion by 2023 \cite{Manokha2019NewEmployees}. Existing commercial solutions include InterGuard, ActivTrak, Time Doctor, Toggl, Activity Monitor, WorkTime Corporate, and Berqun. In a 2007 survey administered by the American Management Association, 73\% of its member companies use technologies to automatically monitor employees \cite{Chory2016OrganizationalResponses}. The legality of workplace surveillance is not debatable. However, employees may have higher psychological stress, and loose trust in the organization if they believe such monitoring does not completely align with organizational justice, which encompasses three dimensions of: distributive, procedural, and interactional justice \cite{Chory2016OrganizationalResponses}. In the case of deploying MPAS internally, it is important to communicate with employees the organizational problems that the tool is helping with solving. Such problems may include: workplace discrimination, sexual harassment, other forms of harassment, suicide, indicators of insider threats, indicators of financial crimes, and so on. Encryption schemes can be used to "mask" the employees' entities, that can only to be unmasked per system's access level.

\section{Conclusion}
Within its limited scope, the paper presented a manual psychoanalysis process snapshot of Steve Jobs case, and proposed the Machine-assisted Psycho-Analysis System (MPAS) for mining psychological artifacts at scale. Unlike most other proposed solutions, this architecture utilizes natural language processing and machine learning at the lower level, allowing faster processing speed, while leveraging evidence-based psychology knowledge and ontological reasoning at the higher level. The use of ontologies provides the much needed explainability and extensibility. For example, reasoning results can be further consumed by graph science and/or model management system for optimizing business processes, understanding team dynamics, predicting insider threats, managing talents, and beyond. It is important to note that issues such as moral disengagement \cite{Erkutlu2019LeaderDisengagement}, and organizational psychopaths' impacts on large scale economy \cite{Boddy2015OrganisationalUpdate} have been under-explored, partially due to the lack of human psychology data.

A technology stack was recommended with references to further resources including starter codes for immediate system development. It is impossible for the paper to present packaged solution due to its limited scope and the needs for customization in building solutions to be used in a particular organization. The paper particularly emphasizes the need to build a custom ontology. It can and should be based on existing evidence-based public researches but it has to be tailored to the organization's people, business, and culture. However, in the case of directing MPAS outward for understanding external audiences, a packaged solution may be possible.

The paper does have a plan for profiling hackers and cyber threat groups using MPAS. The immediate challenge is finding ground truths. Data from cyber criminal investigations is not generally made available to researchers. Even worst, collecting psychological data with regards to hackers' motives, mental needs, and mental issues might have been ignored prior to and even after arrests. It is also challenging to find enough evidence-based research works to support the engineering of an ontology describing hackers psychological states. On a brighter note, predicting the organizational behaviors of hacking groups is possible because their digital breadcrumbs are plenty, and findings can be corroborated by evidences from other domains such as geopolitics, legislation, and economics.


\bibliography{references.bib}

\end{document}